\documentstyle [12pt,eqsecnum,aps,amsfonts] {revtex} 
\input epsf
\newcommand{\beq}{\begin{equation}}
\newcommand{\eeq}{\end{equation}}
\newcommand{\beqs}{\begin{eqnarray}}
\newcommand{\eeqs}{\end{eqnarray}}
\newcommand{\lsim}{\mathrel{\raisebox{-.6ex}{$\stackrel{\textstyle<}{\sim}$}}}
\newcommand{\gsim}{\mathrel{\raisebox{-.6ex}{$\stackrel{\textstyle>}{\sim}$}}}

\begin{document}
\draft

\baselineskip 6.0mm

\title{Some Remarks on Theories with Large Compact Dimensions and 
TeV--Scale Quantum Gravity} 

\vspace{8mm}

\author{
Shmuel Nussinov$^{(a,b)}$ \thanks{email: nussinov@post.tau.ac.il} \and
Robert Shrock$^{(b)}$ \thanks{email: shrock@insti.physics.sunysb.edu}}

\vspace{6mm}

\address{(a) \ Sackler Faculty of Science \\
Tel Aviv University \\
Tel Aviv \\
Ramat Aviv, Israel} 

\address{(b) \ Institute for Theoretical Physics \\
State University of New York \\
Stony Brook, N. Y. 11794-3840, USA}

\maketitle

\vspace{10mm}

\begin{abstract}

We comment on some implications of theories with large compactification radii 
and TeV-scale quantum gravity. These include the behavior of high-energy
gravitational scattering cross sections and consequences for
ultra-high-energy cosmic rays and neutrino scattering, the question of how to
generate naturally light neutrino masses, the issue of quark-lepton
unification, the equivalence principle, and the cosmological constant. 

\end{abstract}

\vspace{16mm}

\pagestyle{empty}
\newpage

\pagestyle{plain}
\pagenumbering{arabic}
\renewcommand{\thefootnote}{\arabic{footnote}}
\setcounter{footnote}{0}

\section{Introduction}

The intriguing idea that fundamental interactions can be understood as
operating in a spacetime of dimension higher than $d=4$ dates back at least to
the work of Kaluza and Klein (KK) \cite{kk}. A number of studies were carried
out subsequently of higher-dimensional field theories, which we shall
generically refer to as Kaluza-Klein theories \cite{rev}.  In a modern context,
Kaluza-Klein theories arise naturally from (super)string theories in the limit
where relevant energies $E$ are much less than the string mass scale $M_s \sim
(\alpha')^{-1/2}$, where $\alpha'$ is the slope parameter.  In both generic
Kaluza-Klein and string theories, there has thus always been the question of
what dynamical mechanism is responsible for compactification and at what
scale(s) $\{R\}$ the extra $n$ spacetime dimensions are compactified, leaving
the observed four spacetime dimensions.  A conventional view has been that the
corresponding compactification mass scale(s) $\{R^{-1}\}$ would be high, $\lsim
M_s$, with $M_s$ being given (in a perturbative analysis) by $M_s \sim g_s
M_{Pl}/\sqrt{8\pi}$, where $M_{Pl} = (\hbar c/G_N)^{1/2} = 1.2 \times 10^{19}$
GeV is the Planck mass, $G_N$ is Newton's constant, and $g_s^2$ is the gauge
coupling at the string scale, of order $g_s^2/(4\pi) \sim 0.04$. However,
recently, there has been considerable interest in the very different and
provocative possibility that some inverse compactification mass scale(s),
$r_c^{-1}$, is (are) much less than the Planck scale \cite{a}-\cite{ddgr}.  A
related feature of this theoretical development is a profound change in the
role of the Newton constant and Planck mass; rather than being fundamental
constants of nature, these become derived quantities, reflecting the change in
spacetime dimensionality, from $d=4$ at large distances, to a higher
dimensionality at distances $r < r_c$ (where for simplicity, we assume
throughout this paper that there is single compactification radius relevant for
gravity) and the resultant change of the gravitational force from 
\beq 
F =
\frac{G_N m_1 m_2}{r^2} = \frac{m_1m_2}{M_{Pl}^2 r^2} \quad {\rm for} \quad r
>> r_c
\label{v4d}
\eeq
to
\beq
F = \frac{G_{N,4+n} m_1 m_2}{r^{2+n}} = 
\frac{m_1 m_2}{\hat M_{4+n}^{n+2}S_{3+n} r^{2+n}} 
\quad {\rm for} \quad r << r_c
\label{v4nd}
\eeq 
where
\beq
S_d= \frac{2\pi^{d/2}}{\Gamma(d/2)}
\label{sd}
\eeq
is the area of the unit sphere in
${\mathbb R}^d$.  Since $r_c$ depends on $n$, it will be denoted as
$r_n$. Setting $M_{4+n}^{n+2}=(2\pi)^n \hat M_{4+n}^{n+2}$, as in
Ref. \cite{add} (motivated by toroidal compactification, in which the volume
of the compactified space is $V_n = (2\pi r_n)^n$) and applying Gauss's law at
$r<< r_n$ and $r >> r_n$, one finds that 
\beq 
M_{Pl}^2 = r_n^n M_{4+n}^{2+n}
\label{mplrel}
\eeq 
i.e., 
\beq 
r_n = M_{4+n}^{-1} \Biggl ( \frac{M_{Pl}}{M_{4+n}} \Biggr )^{2/n} = 
(2.0 \times 10^{-17} \ {\rm cm} \ ) \Biggl ( \frac{ 1 \ {\rm TeV}}{M_{4+n}} 
\Biggr )\Biggl ( \frac{M_{Pl}}{M_{4+n}} \Biggr )^{2/n}
\label{rnvalue}
\eeq 
Assuming that the higher-dimensional theory at short distances is 
a string theory, one expects that the fundamental string scale $M_s$ and Planck
mass $M_{4+n}$ are not too different (a perturbative expectation is that
$M_s \sim g_s M_{4+n}$).  Thus, a 
compactification radius such that $r_n^{-1} << M_{Pl}$ corresponds to a 
short-distance Planck scale and string mass $M_s$ which are also $<< M_{Pl}$. 

It is a striking fact that there is an vast extrapolation of 33 orders of
magnitude between the smallest scale of O(1) cm to which Newton's law has been
tested \cite{w,gravexp} and the scale that has conventionally been regarded as
being characteristic of quantum gravity, namely the Planck length, $L_{Pl} =
\hbar/(M_{Pl}c) \sim 10^{-33}$ cm, and it is not at all implausible that new
phenomena could occur in these 33 decades that would significantly modify the
nature of gravity.  It therefore instructive to explore how drastically one can
change the conventional scenario in which both gauge and gravitational
interactions occur in four-dimensional spacetime up to energies comparable to
the Planck mass.  Of course if one had a truly fundamental theory of
everything, it would predict and explain the scale(s) of compactification and
the changes in dimensionality that occur.  Here we shall take a
phenomenological attitude of considering various values of $n$.  From
eq. (\ref{mplrel}) it follows that for fixed $M_{4+n}$, $r_n$ is a
monotonically decreasing function of $n$.  (From eq. (\ref{rnvalue}), one sees
that in the formal limit $n \to \infty$, $r_n$ approaches $M_{4+n}^{-1}$ from
above; in a string theory context, the values of $n$ up to 6 are of interest
since this corresponds to spacetime dimensions up to 10 at short distances.) 
Consequently, the strongest
challenge to the conventional paradigm is obtained for the smallest values of
$M_{4+n}$ and $n$.  From this point of view, one is therefore motivated to
consider values of $M_{4+n}$ as low as the 1-10 TeV region.  For such values,
the case $n=1$ would yield a compactification radii larger than the solar
system, and hence is clearly excluded by existing measurements of gravity and
tests of Newton's law. For $M_{4+n} \sim 30$ TeV (which in fact is a lower
bound \cite{add2}), the case $n=2$ yields a compactification radius 
\beq
r_{n=2} \equiv r_2 \simeq 2.7 \ {\rm microns}, \ i.e., \quad r_2^{-1} \sim 0.07
\ {\rm eV} \quad {\rm for} \quad M_{4+n}=M_6=30 \ {\rm TeV}
\label{r2}
\eeq 
As $r$ decreases below this scale, the gravitational force changes from a
$1/r^2$ to a $1/r^4$ behavior.  Currently planned experiments plan to probe
gravity somewhat below the present limit of O(1) cm \cite{gravexp}.  We shall
concentrate on the case $n=2$ because, among the allowed values $n \ge 2$, it
yields (for a given $M_{4+n}$) the largest value of the compactification radius
and hence the strongest contrast to the conventional paradigm.

Another reason for considering theories with very low string scales not too far
above the electroweak scale, $M_{ew} = 2^{-1/4}G_F^{-1/2} = 250$ GeV, is that
this essentially removes the old hierarchy problem, i.e., the problem of
preventing the Higgs mass from getting large radiative corrections that would
naturally raise it to the GUT or conventional string ($\sim$ Planck) scale.
One must acknowledge that a new hierarchy appears, namely the large ratio
between the compactification mass $r_n^{-1}$ and the string scale.  For $n=2$,
with $M_s \sim 1$ TeV, and $r_2^{-1}$ as given in eq. (\ref{r2}), this ratio is
$\sim (1 \ {\rm TeV})/(10^{-1})$ eV $= 10^{13}$, which is almost as large as
the old hierarchy $M_{GUT}/M_{ew}$ or $M_{Pl}/M_{ew}$.  Obviously,
supersymmetry cannot be used to stabilize this new hierarchy since it is broken
at a scale of at least the electroweak level; some ideas for how this
stabilization might occur have been discussed recently \cite{stab,ddgr}.  Note
that if, indeed, the string scale is as low as $\sim 1$ TeV, so that the
conventional hierarchy is absent, one motivation for supersymmetry would be
removed, although its original motivation -- to avoid tachyons in string theory
-- would still be present, given that the quantum theory of gravity is assumed 
to be a string theory.

In a theory with extra dimensions compactified at a scale $r_n$, it would
naively seem that for distances much less than the compactification scale, all
of the fields would depend on the coordinates of the higher-dimensional space.
Of course, if this were true, then low values of $n$ including $n=2$ would
clearly be ruled out since, among other things, QCD and electroweak
interactions have been well measured up to energies of order $10^2$ GeV
(lengths down to $10^{-16}$ cm), and the data shows that these interactions
take place in a spacetime of dimension 4.  Hence, to avoid the danger of a
contradiction with experiment, one is led to require that the known fermions
and gauge fields be confined to four-dimensional spacetime at least down to
distances of about (1 \ TeV)$^{-1} \sim 10^{-17}$ cm.  Several possible
mechanisms for this dimensional confinement of standard model fields have been
suggested \cite{w,add,ddg}.  A particularly appealing mechanism is present in
modern string theories with Dirichlet $p$-branes D$_p$ (commonly denoted 
D-branes \cite{dbrane}); see, e.g., \cite{w,ddg,st,sundrum,dbranerev} and
references therein.  Calculations of scattering processes involving D-branes
suggest that when probed at high energy, these exhibit a thickness \cite{delta}
\beq 
\delta \sim M_s^{-1}
\label{delta}
\eeq 
Specifically, as $r$ decreases past $r_n$, gravity would feel the extra
$n$ dimensions, but the usual gauge and matter fields would be confined to a
$p=3$ D$_p$-brane sweeping out the usual Minkowski 4-dimensional spacetime.
The fact that the gravitons do propagate in all $4+n$ dimensions is responsible
for the change in the gravitational force law from $1/r^2$ to $1/r^{2+n}$ at
distance scales below the compactification scale $r_n$. There are several
specific scenarios of this type.  One type of example features a type-I string
theory with 5-branes and 9-branes sweeping out 6-dimensional and 10-dimensional
spacetime volumes, respectively, and each having noncompact 4-dimensional
Minkowski submanifolds \cite{aadd,st}.  As $r$ decreases below $r_c$, gravitons
(the closed-strings) change from propagating in four dimensions to propagating
in six dimensions, so that $n=2$ and $r_c=r_2$, but gauge and matter fields
(corresponding to open string states) continue to reside on 3-branes in the
5-branes.  As $r$ decreases through an additional compactification scale
slightly above $M_s^{-1}$, the gauge and matter fields extend to a 9-brane
sweeping out the full 10-dimensional spacetime.  In this region the gauge
couplings run rapidly, since they have dimensions; studies of how these
couplings might unify at the TeV scale (using several different specific
models) include Refs. \cite{ddg,st}.  If, as in the standard model and
supersymmetric generalizations thereof, the fermions gain their masses from
Yukawa couplings, then these also run rapidly for the same reason.

Given the provocative new features of these proposed models with large compact
dimensions and TeV scale strings, there is strong motivation for immediate
phenomenological studies to assess their experimental viability, and these have
been initiated in a number of works, e.g., Refs. \cite{add}-\cite{sundrum}.  
Important
issues that have been studied include the above-mentioned constraints due to
experimental gravity tests, and also proton decay, possible contributions to
flavor-changing neutral currents and precision electroweak observables, effects
mimicking compositeness and changes in scattering processes measured in current
$e^+e^-$ and $\bar p p$ collider experiments, rare decays, and astrophysical
and cosmological effects.  One serious concern is that the contributions of
Kaluza-Klein modes to the mean mass energy would overclose the universe;
however, it has been argued that the theory can evade this problem \cite{add2}.
Another severe constraint arises from the effects of KK-graviton emission on
cooling of supernovae.  This has been used to infer the lower bound \cite{add2}
\beq
 M_{4+n} \gsim 10^{(15-4.5n)/(n+2)} \ \ {\rm TeV}
\label{mbound}
\eeq
i.e., for the case of main interest here, $n=2$, 
\beq
M_6 \gsim 30 \ \ {\rm TeV}
\label{m6bound}
\eeq 
It has been argued that this may still be consistent with a fundamental
string scale $M_s$ of O(1) TeV \cite{add2}.  Implications for the cosmological
constant have also been discussed (e.g., Refs. \cite{add2,sundrum,stab,ddgr}). 
Effects on dispersion of
light travelling over cosmological distances may also yield serious constraints
\cite{lightdisp}.  The problem of stablizing the new hierarchy $M_s r_c >> 1$
has been addressed in several papers, including Refs. \cite{stab,ddgr}.

\vspace{4mm}

In this paper we shall remark on some other phenomenological implications of
these theories with large compact dimensions and TeV--scale strings.  In 
section 2 we review how the exchange of KK modes of gravitons can produce 
relatively large effects.  We then give some estimates of their 
effects on high-energy scattering cross sections.  In section 3 we address the
problem of obtaining light neutrino masses in the absence of the conventional
methods (seesaw mechanism and higher-dimension operators).  Sections 4 and 5
contain some discussion of the equivalence principle and the cosmological
constant. 

\section{Graviton+KK Exchanges and High-Energy Behavior of Cross Sections}

In the theories under consideration here, there are several relevent ranges for
the center-of-mass energy $\sqrt{s}$ of a given process: (i) the extreme
low-energy region, $\sqrt{s} < r_2^{-1}$; (ii) the large range $r_2^{-1} <
\sqrt{s} < M_s$ which includes energies up to the TeV scale at its upper end:
and (iii) the range of energies above the string scale, $\sqrt{s} > M_s \sim 1$
TeV.  In this section, by ``high-energy'' behavior of cross sections, we shall
mean intervals (ii) and (iii).  Let us denote
the momentum of a graviton as $k = k_L = (\{k_\lambda\},k_1,...,k_n)$, where
the usual spacetime Lorentz index $\lambda=0,1,2,3$.  Because of the
compactification, the extra $n$ components of the graviton momenta are
quantized. With the simplifying assumption that the compactification radii of
all of the extra $n$ dimensions are the same, one has, for toroidal
compactification, with the circumferences \beq L_i = 2 \pi r_n \equiv L_n \ ,
\quad i = 1,...,n
\label{li}
\eeq
the quantization of KK momenta
\beq 
k_i = \frac{2\pi \ell_i}{L_n} = \frac{\ell_i}{r_n} \ , \quad i=1,...,n
\label{kkquantization}
\eeq
To an observer in the usual four-dimensional spacetime, the above graviton 
would thus appear to be a massive Kaluza-Klein (KK) state with mass
\beq
\mu_{\ell_1,...,\ell_n} = \Bigl ( \sum_{i=1}^n \ell_i^2 \Bigr )^{1/2} 
r_n^{-1}
\label{mukk}
\eeq 
All of these KK states have the same Lorentz structure as the graviton as
regards their couplings to other particles.  Since the gravitons propagate in
the full $(4+n)$--dimensional spacetime, their self-interactions interactions
must conserve not only the ordinary 4-momenta, but also the KK momentum
components.  That is, if one envisions a scattering process involving $N$
gravitons with momenta $k_L^{(1)},..., k_L^{(N)}$ (directed into the vertex,
say), then $\sum_{j=1}^N k_L^{(j)}=0$, so that $\sum_{j=1}^N k_\lambda^{(j)}=0$
for the usual spacetime components $\lambda=0,1,2,3$, and also $\sum_{j=1}^N
k_i^{(j)}=0$ for $i=1,...,n$, whence $\sum_{j=1}^N \ell_i^{(j)}=0$ for
$i=1,...,n$.  However, since other particles are assumed to be confined to the
thin membrane of thickness $\delta \sim M_s^{-1}$, which breaks translational
invariance in the extra $n$ dimensions, they do not have well-defined KK
momenta in these extra dimensions.  Therefore the interactions of gravitons
with such particles do not, in general, conserve the KK momentum components,
$\{\ell_i/r_n \}$, $i=1,...,n$ -- at least so long as these are smaller than
$M_s$, the ultimate cutoff scale in the field theory, at which it goes over
into a string theory, i.e., so long as \beq \frac{\ell_i}{r_n} \lsim M_s
\label{cutoff}
\eeq
Defining the graviton field as
\beq
g_{MN} = \eta_{MN} + \frac{h_{MN}}{(M_{4+n}^{2+n})^{1/2}}
\label{hmn}
\eeq 
(where $\eta_{\mu \nu}$ is the usual flat-space metric tensor, with signature
which we take as $(+,-,-,-)$, and the details of $\eta_{MN}$ in the extra
dimensions depend on the nature of the compactified manifold), 
the resulting interactions of the gravitons with the usual gauge and
matter fields on the 3-brane are given by $T_{\mu \nu}(x) h_{\mu \nu}(x)/
(M_{4+n}^{2+n})^{1/2}$ for $x$ restricted to lie on this 3-brane.
Equivalently, one can treat the graviton-KK emission in a four-dimensional
framework, where the coupling is $1/M_{Pl}$ in an amplitude; the rate is then
proportional to $1/M_{Pl}^2$ times a factor reflecting the multiplicity of
KK-graviton emission.  Since this factor is $\sim (s^{1/2}r_n)^n$, where
$s^{1/2}$ is the center-of-mass energy available for graviton-KK emission, when
one subsitutes the expression for $r_n$ from eq. (\ref{mplrel}), the factor of
$1/M_{Pl}^2$ is exactly cancelled, and the final product is 
$s^{n/2}/M_{4+n}^{n+2}$, as one would obtain directly from eq. (\ref{hmn})
\cite{add}.  Thus, from a four-dimensional viewpoint, although the
KK-gravitons are coupled extremely weakly, this is compensated by their
very large multiplicity, so that their net effect involves in the denominator 
a mass scale in the range of 10 TeV instead of $M_{Pl}$.

Let us study the implications of this further. In considering the exchange of
gravitons, and in particular, their KK components, in some process, one should
formally consider all $\ell_i \in \mathbb Z$ for each $i=1,...,n$ and sum over
all of these exchanges.  In the theories of interest here, as $\sqrt{s}$
becomes comparable to the string scale, $M_s$, one changes over from a field
theory (with effects of D-branes included) to a fully stringlike picture, so
that $M_s$ serves as an upper cutoff to what is really the low-energy effective
field theory with which we work.  Accordingly, we shall impose an upper cutoff
\beq 
\ell_i < \ell_{max} = M_s r_n
\label{ellmax}
\eeq
on the sums over KK modes, which thus run over the range
\beq
\ell_i = 0, \pm 1, ... , \pm \ell_{max} \quad {\rm for} \quad i=1,...,n
\label{kkrange}
\eeq
The value of $\ell_{max}=M_sr_n$ is very large; for example, for the case of 
primary interest here, $n=2$, for $M_s \sim 1$ TeV, one has 
$\ell_{max} = M_s r_2 \sim 10^{16}$ while for $n=3,4$, 
$\ell_{max} = M_s r_n \sim 10^{11}$ and $10^8$, respectively. 
When one is interested in the effect of
the exchange of these KK components of gravitons on the static gravitational
potentional generated between two test masses at a distance $r$, the
resultant contribution of the higher KK modes is suppressed by a 
Yukawa-type factor: 
\beq
\frac{V}{m_1 m_2} = \frac{G_N}{r} \sum_{\ell_1, ...\ell_n}
e^{-\mu r} = \frac{G_N}{r}\sum_{\ell_1,...,\ell_n} \exp \biggl 
[ -(\sum_{j=1}^n \ell_j^2)^{1/2} \biggr ]  
\label{yuk}
\eeq
Formally, the summation is over each $\ell_i \in \mathbb Z$, but the actual
contribution depends on the value of $r$ relative to $r_n$ since only the 
number 
\beq
\nu_{eff} \sim \biggl (\frac{r_n}{r} \biggr )^n
\label{neff}
\eeq
of KK modes which do not suffer exponential suppression in eq. (\ref{yuk})
contribute significantly.  Thus, when $r >> r_n$, 
only the term with $(\ell_1,...,\ell_n)=(0,...0)$ contributes and one recovers
the usual Newton law $V \propto 1/r$, but as $r$ decreases through $r_n$, more
KK modes contribute, and finally, for $r << r_n$, all of the KK modes 
up to $M_sr$ contribute, giving rise to the crossover in the behavior of the 
gravitational interaction $V \to 1/r^{1+n}$.

The consideration of the exchange or production of these KK components of
gravitons in various processes, while not exhausting the full implications of
the new theory, can serve as a general guideline for studying the possible
novel phenomenological implications.  Note also that unlike other possible
signatures such as missing energy, apparent compositeness, etc. which could
originate from other modifications/extensions of the standard model, the
multi-KK exchange and some of its consequences are unique to the present class
of models. 

Already at present, cosmic ray interactions have provided us with $pp$
collisions having lab and center of mass energies up to $\sim 3 \times 10^8$
TeV and $\sim 10^3$ TeV, respectively \cite{uhecr}, far above the assumed
string scale $M_s$, the ultimate cutoff of the field theory.  It has been
heuristically argued that since the bulk of the hadronic collisions are soft,
the new TeV region physics will not be strongly manifest in these collisions
\cite{add2}.  One of our purposes in this paper will be to examine this issue
and the more general question of the impact of graviton+KK exchange(s) in
quasi-diffractive processes which are defined here to be $2 \to 2$ scattering
processes with cm mass energy squared $s >> M_s^2$ but $Q^2 < M_s^2$ (or even
$Q^2 << M_s^2$).   Using the strong lower bound (\ref{mbound}), which forces
$M_{4+n}$ to be substantially above the electroweak scale, we obtain results
that confirm the arguments of Refs. \cite{add2}.

Before proceeding, a comment is in order: in a generic Kaluza-Klein theory,
there would also be KK excitations of usual standard-model gauge, fermion, and
Higgs fields. However, in the type of theory that we consider here, since the
compactification radius $r_2$ is so large, and hence, as noted above, 
the standard-model fields must be confined to 
usual 4-dimensional spacetime even for $r < r_c$, down to distances $r
\sim M_s^{-1}$ where a fully stringlike picture emerges, it follows that 
standard-model fields do not have KK excitations in the energy range $\sqrt{s}
\lsim M_s$.

Let us consider the scattering of two usual standard-model particles, indicated
by their momenta, $p_1 + p_2 \to p_1' + p_2'$.  These could be leptons and/or
the quark/gluon constituents of energetic protons.  As stated above, we
concentrate on the case of $n=2$ extra dimensions whose compactification radius
is given by (\ref{r2}).  We define $P=p_1+p_2=p_1'+p_2'$ and
$q=p_1-p_1'=p_2'-p_2$.  The center of mass energy squared is then given by 
\beq
s=P^2 = \sum_{M,N} P_M P_N g^{MN} =
\sum_{\mu,\nu=0,1,2,3} P_\mu P_\nu g^{\mu \nu}
\label{s}
\eeq
and the momentum transfer squared $t=q^2=-Q^2$ by 
\beq
t=q^2=\sum_{M,N} q_M q_N g^{MN} 
= \sum_{\mu,\nu=0,1,2,3} q_\mu q_\nu g^{\mu \nu}
\label{t}
\eeq
where we have used the fact that the usual standard-model particles are
confined to four-dimensional spacetime.  The third kinematic invariant is 
$u=(p_1-p_2')^2$ satisfying the relation that $s+t+u$ is equal to the sums of
the squared masses of the colliding particles.  A vector gauge 
boson $V$ ($= \gamma$, $Z^0$, $W$, or gluon) exchanged between the particles 
yields a t-channel Born amplitude
which, for 
\beq
s >> m_V^2 \ , \quad s >> t \ , \quad {\rm and} \quad s >> m_i^2
\label{scatcondition}
\eeq
(whence $u \sim -s$) is of the form 
\beq
A_{1v} \simeq \frac{g_1 g_2 (p_1+p_1')_\mu (p_2 + p_2')_\nu g^{\mu \nu}}
{t-m_V^2} \simeq \frac{g_1 g_2 (s-u)}{t - m_V^2} = \frac{2g_1 g_2 s}{t-m_V^2}
\label{vecbosamp}
\eeq
where $g_1, g_2$ are appropriate gauge coupling constants, 
the $(p_j+p_j')_\mu$ factors arise from the gauge-fermion couplings, 
and $m_V$ is the mass of the exchanged vector boson.  We neglect the 
contributions from the $q_\mu q_\nu/m_W^2$ term in the $W$ propagator 
numerator which give negligible contributions, and the respective 
$q_\mu q_\nu/m_Z^2$ and (gauge-dependent) $q_\mu q_\nu/q^2$ term in the $Z$ 
and $\gamma$ propagator numerators, which give zero contributions 
because the $Z$ and $\gamma$ have diagonal, current-conserving, couplings. 

We next consider the graviton exchange amplitude and begin with the usual
four-dimensional massless graviton exchange.  It has long been recognized that
the (perturbatively calculated) scattering amplitudes involving the exchange of
a particle with spin $J$ in the $t$ channel grow like $s^J$; indeed, this was
one of the motivations for the development of Regge theory, in which the spins
of the exchanged particles were effectively made into a variable $\alpha(t)$
and in which the resultant energy-dependence of the amplitude was much softer
(e.g., \cite{cseden}). Since gravity couples to the (conserved, traceless,
symmetric) energy-momentum tensor, the amplitude for the exchange of one
graviton between the same particles, with the same kinematic condition
(\ref{scatcondition}) yields an amplitude 
\beqs 
A_{1g} & \simeq & G_N\Bigl [
(p_1+p_1')_\mu (p_1+p_1')_\nu - (p_1+p_1')^2 g_{\mu \nu} \Bigr ] \times \cr & &
\Bigl [(p_2+p_2')_\rho (p_2+p_2')_\sigma-(p_2+p_2')^2 g_{\rho \sigma} \Bigr ]
\frac{(g^{\mu \rho} g^{\nu \sigma} + perms.)}{t} \cr & \propto & \frac{G_N
s^2}{t}
\label{gravamp}
\eeqs 
where the conserved 
nature of the graviton-fermion couplings means that terms involving momenta
$q$ in the numerator of the graviton propagator do not contribute. 
The $s^2$ behavior in eq. (\ref{gravamp}) conforms to the expected $s^J$
behavior for spin-$J$ exchange.  Comparing with eq. (\ref{vecbosamp}), we see
that 
\beq 
\frac{A_{1g}}{A_{1v}} \simeq \frac{s/M_{Pl}^2}{g_1g_2}
\label{gvcomparison}
\eeq 
This ratio becomes of order unity for $\sqrt{s} \sim M_{Pl}$, i.e., 
ordinary four-dimensional quantum gravity becomes strong only at superplanckian
energies. 

However, the contribution of the many KK components of gravitons changes this 
picture significantly.  Reverting to the case of general $n$, we observe that, 
from eqs. (\ref{cutoff}) and (\ref{kkrange}), this contribution has the 
effect of replacing the propagator $1/Q^2$ by 
\beq
\frac{1}{Q^2} \rightarrow \sum_{\ell_1} \cdot \cdot \cdot 
\sum_{\ell_n} \frac{1}{-q_0^2 + |{\bf q}|^2 + 
\sum_{i=1}^n (\ell_i/r_n)^2}
\label{kkeffect}
\eeq where the range of the indices $\ell_i$, $i=1,..n$ in the summation is
given by eq. (\ref{kkrange}).  To obtain an approximate value for this
expression, we note that there is a large range of each KK momentum squared,
$(\ell_i/r_n)^2$, from 0 to $M_s^2$, 
and over much of this range, it is $>> Q^2$
for typical $Q^2$ values of $Q^2 \sim \Lambda_{QCD}^2 \sim (150 \ {\rm MeV})^2$
that give the dominant contribute to hadronic cross sections.  Similar comments
can be made about cross sections for other types of reactions.  Therefore, we
consider the approximation of keeping only the KK graviton modes in
eq. (\ref{kkeffect}) satisfying 
\beq 
\sum_{i=0}^n (\ell_i/r_n)^2 >> Q^2
\label{kksumcondition}
\eeq 
For $n \ge 3$, we then drop the $Q^2$ term in the denominator of
(\ref{kkeffect}).  Further, recalling that $\ell_{max}$ is very large, it is 
reasonable to approximate the summations by integrations, yielding, 
\beq
\frac{1}{Q^2} \rightarrow 
r_n^2 \int_{-\ell_{max}}^{\ell_{max}} \cdot \cdot \cdot 
\int_{\ell_{max}}^{\ell_{max}} 
\frac{d^n \ell}{\ell^2} = \frac{r_n^2 S_n (M_s r_n)^{n-2}}{n-2} \quad 
{\rm for} \quad n \ge 2 
\label{integral}
\eeq
where $S_n$ was given in (\ref{sd}). For $n=2$, we must retain the $Q^2$ 
term in the denominator of (\ref{kkeffect}), and, again approximating the
summations by integrations, we obtain
\beq
\frac{1}{Q^2} \rightarrow 
\pi r_2^2 \ln \biggl ( 1 + \frac{M_s^2}{Q^2} \biggr ) \quad {\rm for} \quad 
n=2
\label{integralneq2}
\eeq
Next, substituting eq. (\ref{kkeffect}) in eq. (\ref{gravamp}), using the 
above approximations, and inserting eq. (\ref{rnvalue}) for $r_n$, we find 
for $n =2$ \beq
\tilde A_{grav.} = \sum_{\ell_1, ..., \ell_n} A_{1g+ KK} = 
\frac{\pi s^2}{M_6^4}\ln \Bigl ( 1 + \frac{M_s^2}{Q^2} \Bigr )
\label{atildeneq2}
\eeq
and for $n \ge 3$, 
\beq
\tilde A_{grav.} = \sum_{\ell_1, ..., \ell_n} A_{1g+ KK} = 
\frac{S_n s^2}{M_{4+n}^{4}} \biggl ( \frac{M_s}{M_{4+n}} \biggr )^{n-2} 
\label{atildengt2}
\eeq
As one might expect, the amplitude $\tilde A_{grav.}$ becomes large once $s$
significantly exceeds $M_{4+n}^2$. 
In particular, we find, in this new Born
approximation of one graviton + KK exchange, a differential cross section
which, for $n=2$, is 
\beq
\frac{d\sigma}{dt} = \frac{1}{16 \pi s^2} |\tilde A_{grav.}|^2 \simeq 
\frac{\pi s^2}{16 M_6^8} \ln^2 \Bigl ( 1 + \frac{M_s^2}{Q^2} \Bigr )
\label{dsigmadtneq2}
\eeq
and, for $n \ge 3$, is 
\beq
\frac{d\sigma}{dt} \simeq \frac{S_n^2 s^2}{16 \pi M_{4+n}^8} \biggl (
\frac{M_s}{M_{4+n}} \biggr )^{2(n-2)}
\label{dsigmadtnge3}
\eeq
In the following we shall restrict ourselves to momentum transfers that are
soft on the scale of $M_s$, i.e., to 
\beq 
Q^2 \lsim \xi M_s^2
\label{tlessthanms2}
\eeq 
where we take $\xi \lsim 10^{-2}$.  We also apply a similar softness cutoff
to the allowed KK graviton exchanges by having the sums in eq. (\ref{kkeffect})
extend not over the range $0 \le |\ell_i| \le M_sr_n$ but instead only over the
range $0 \le |\ell_i| \le \xi^{1/2} M_sr_n$. 
This will simply introduce an extra $\xi^{(n-2)/2}$ factor in eq. 
(\ref{integral}) and, for $n=2$ will remove the
logarithmic enhancement factors in eq. (\ref{integralneq2}) and 
(\ref{dsigmadtneq2}).
Integrating over the region $0 \le -t \le \xi M_s^2$, we find the total cross 
sections, for $n=2$, 
\beq
\sigma = \int \frac{d\sigma}{dt} dt \ge \int_{-\xi M_s^2}^0 \frac{d\sigma}{dt} 
dt \simeq \frac{\pi \xi s^2 M_s^2 }{16 M_6^8}  \simeq \xi \biggl ( 
\frac{s}{1 \ {\rm TeV}^2} \biggr )^2
\biggl ( \frac{1 \ {\rm TeV} }{M_6} \biggr )^6 \biggl (
\frac{M_s}{M_6} \biggr )^2 \times 10^{-34}  \ \ {\rm cm}^2
\label{sigmaval}
\eeq
and, for $n \ge 3$, 
\beq
\sigma \simeq \frac{\xi S_n^2 s^2 M_s^2}{16 \pi M_{4+n}^8} 
\biggl (\frac{M_s}{M_{4+n}}\biggr )^{2(n-2)} 
\simeq \xi S_n^2 \biggl ( \frac{s}{1 \ {\rm TeV}^2} \biggr )^2 
\biggl ( \frac{1 \ {\rm TeV}}{M_{4+n}} \biggr )^8 
\biggl ( \frac{M_s}{M_{4+n}} \biggr )^{2(n-2)} \ \times 10^{-35} \ 
{\rm cm}^2
\label{sigmange3}
\eeq 
As noted before, this $\sigma \propto s^2$ growth is characteristic of a
process involving the exchange of a spin-2 particle (see also
Refs. \cite{soldate,thooft}).  One can, in addition, 
consider the contributions of higher-order graviton+KK exchanges, but it is a 
challenging problem to calculate these (see further below). 

  We next consider a different approach which yields, by construction, a
unitarized amplitude and cross section.  As elsewhere in this paper, we focus
on the case $n=2$.  The unitarization procedure starts by performing a partial
wave decomposition of the original, non-unitarized Born amplitude.  The
resulting partial waves are (in general, away from possible resonances)
monotonically decreasing with $\ell$.  For our estimate, let us assume that for
$\ell \le \ell_0(s)$, all of the partial waves exceed the unitarity bound
$a_\ell(s) \le 1$ and for all $\ell > \ell_0(s)$ they obey this bound.  The
method is then to set all $a_\ell(s)$ for $ \ell < \ell_0(s)$ equal to unity.
In the present context of small momentum transfer, $s >> -t$, and hence small
center-of-mass scattering angle $\theta \equiv \theta_{cm}$ (recall
$\theta_{cm}^2 \sim -4t/s$ in this limit), we can make use of the eikonal or
related peripheral approximation (for a review, see, e.g., \cite{cseden}).  We
introduce the notation ${\bf k} \equiv {\bf p_1}_{cm}$ and ${\bf k'} \equiv
{\bf p_1'}$, and note that $k = |{\bf k}| = |{\bf k'}| \simeq \sqrt{s}/2$ in
the limit that we are considering. The impact parameter ${\bf b}$ satisfies
${\bf b} \ \bot \ \hat {\bf k}$, so ${\bf q} \cdot {\bf b} = -{\bf k'} \cdot
{\bf b} \simeq kb\theta \cos\phi$, where $\phi$ is the azimuthal angle relative
to $\hat {\bf k}$.  Denoting $|{\bf b}| = b$, we recall that $\ell=|{\bf q}|b$.
We can thus write an integral transform defining the eikonal amplitude $\tilde
a(s,b)$: \beqs A(s,t) & = & s \int \frac{d^2 {\bf b}}{2\pi} \ e^{i {\bf q}
\cdot {\bf b}} \ \tilde a(s,b) \\ & & \\ & = & s \int_0^\infty b db
\int_0^{2\pi} d\phi \ e^{-ikb\theta \cos\phi} \ \tilde a(s,b) \\ & & \\ & = & s
\int_0^\infty b db \ J_0(kb\theta) \ \tilde a(s,b)
\label{ast}
\eeqs
The inverse relation is then, formally, 
\beq
\tilde a(s,b) = \frac{1}{s}\int \frac{d^2 {\bf q}}{2\pi} e^{i{\bf q} \cdot 
{\bf b}} A(s,t) 
\label{inverse}
\eeq
where it is understood that $Q^2 \simeq |{\bf q}|^2 < \xi M_s^2 << M_s^2$
because of the quasi-diffractive kinematic conditions assumed for our
estimate. In accordance with our unitarization, we set the eikonal function 
$\tilde a(s,b)$ equal to unity for $b \le b_0$, where 
$b_0=\ell_0/|{\bf q}|$.  A rough estimate of the total cross section is then 
\beq 
\sigma \simeq \pi b_0^2
\label{sigmabsq}
\eeq
Substituting (\ref{atildeneq2}) into the inverse transform 
(\ref{atildeneq2}), we obtain the dimensional estimate 
\beq
\tilde a(s,b) \sim const. \times \frac{s}{M_6^4 b^2 }
\label{asb}
\eeq
This function $\tilde a(s,b)$ decreases monotonically as a function of $b$, 
and 
\beq
\tilde a(s,b)=1 \quad {\rm for} \quad b_0^2 \sim  const. \times 
\frac{s}{M_6^4}
\label{asb1}
\eeq
Hence, from eq. (\ref{sigmabsq}) we have the unitarized estimate 
\beq
\sigma \propto \frac{s}{M_6^4}
\label{sigmaeikonal}
\eeq 
Evidently, the behavior $\sigma \propto s$ is a less rapid growth than is
suggested by the analysis of 1-graviton exchange above, which yielded the
behavior $\sigma \propto s^2$.  We consider the eikonal method to
be more reliable because, by construction, it produces a unitarized amplitude,
whereas the 1-graviton+KK exchange calculation (and the additions of multiple
graviton+KK exchange) does not automatically do this. 

Another way of seeing this result is to consider two particles of energy $W/2$,
each moving in opposite directions in the center of mass frame, so that the
compound system has energy $W$ ($=\sqrt{s}$).  Now recall that
the Schwarzschild radius of a black hole of mass $m$ is $R_{Sch.} = 2G_N m =
2m/M_{Pl}^2$.  Therefore, if the impact parameter (size) of the system is
smaller than $ \sim 2W/M_{Pl}^2$, we might expect it to collapse to a
black hole.  The critical impact parameter $b_0 \simeq 2W/M_{Pl}^2$ may thus
serve as an effective interaction range.  Note that this argument applies for
both elastic scattering and inelastic scattering involving multiparticle 
production.  This then leads to the estimate for the total 
cross section $\sigma \simeq \pi b_0^2 \simeq (4 \pi s)/M_{Pl}^4$.  
Making the replacement $M_{Pl} \to M_{4+n}$ appropriate for the 
models considered here (provided that $b_0 << r_n$), we obtain for the total
cross section 
\beq
\sigma \simeq \frac{4 \pi s}{M_{4+n}^4}
\label{sigmafinaln}
\eeq
In particular, for the case $n=2$ of main interest here, 
\beq
\sigma \simeq \frac{4 \pi s}{M_6^4} \simeq 5 \biggl ( \frac{s}{1 \ {\rm TeV^2}}
\biggr )\biggl ( \frac{1 \ {\rm TeV}}{M_6}\biggr )^4 \ \times 10^{-33} \ {\rm
cm^2} 
\label{sigmafinal}
\eeq
which exhibits a $\sigma \propto s$ behavior, in agreement with 
(\ref{sigmaeikonal}).  Given the lower bound $M_6 \gsim 30$ TeV in eq. 
(\ref{m6bound}), we have 
\beq
\sigma \lsim 0.6 \biggl ( \frac{s}{1 \ {\rm TeV^2}}\biggr ) \times 10^{-38} \ 
{\rm cm}^2 
\label{sigmafinalbound}
\eeq
We shall discuss the phenomenological implications of this below. 

In passing, we note that the behavior $\sigma \simeq s$ of eqs. 
(\ref{sigmaeikonal}) and (\ref{sigmafinal})
may be viewed as the analogue of the Froissart bound \cite{frois} 
\beq
\sigma \le \frac{4\pi}{\mu^2}\ln^2 \Bigl ( \frac{s}{s_0} \Bigr )
\label{frois}
\eeq 
in ordinary hadronic physics.  Since the quantity $\mu^2$ that appears in
eq. (\ref{frois}) is the squared mass of the lightest particle exchanged in the
$t$ channel, which for the present case is the massless graviton, the original
Froissart bound (\cite{frois}) is clearly not applicable here.  In the
following we will adopt the conservative estimate (\ref{sigmafinal}). 

We have seen that the behavior $\sigma \propto s$ does not conflict with any
unitarity bounds; in fact, we have incorporated these bounds for our estimate
of the total cross section.  Our arguments above based on the eikonal model and
black-hole considerations are rather general and do not depend on calculating
specific field-theory diagrams.  This is important, since in the energy region
$s > M_s^2$, there are stringy effects.  Although
it is not necessary for our arguments, we briefly comment on these.  Once $W >
M_s$, the massive string states consisting of a tower $M^2 = rM_s^2$, where $r=
1,2,...$ can be excited.  These have exponentially rising degeneracy $\rho_r
\sim \exp(\sqrt{r})$.  A description in terms of $s$-channel string state
production would be challenging, expecially given the necessity of carrying out
fully the multi-loop string unitarization program to all orders in order to
avoid poles on the real $s$ axis and obtain physical behavior involving
finite-wide resonances in $A(s,t)$ \cite{reggecuts,sv}.  However, precisely
because of the basic feature of string models (which actually predates
strings), namely the Horn-Schmid duality \cite{duality}, we need not compute
$A_{2 \to 2}(s,t)$ by summing the contributions of $s$-channel resonances.
This specific duality implies that we can equally well represent $A_{2 \to
2}(s,t)$ by summing over $t$-channel contributions.  In particular, for the
region of interest here, namely $s >> M_s^2$, $-t << M_s^2$, the second,
asymptotic series description is much easier to compute, as it consists of only
the contribution from the leading Regge trajectories.  Hence, 
\beq 
A(s,t)
\simeq \beta(t)s^{\alpha(t)} \quad {\rm for} \quad s \to \infty \quad {\rm and}
\quad \frac{-t}{s} << 1
\label{astregge}
\eeq
In the present case we know what the leading Regge trajectory is, namely the
graviton, with intercept $\alpha(t=0)=J=2$ and slope 
\beq
\alpha_G(t) = 2 - \alpha' t \quad {\rm with} \quad \alpha'=\frac{1}{4 \pi T} =
\frac{1}{4 \pi M_s^2}
\label{alphag}
\eeq
where $T$ is the string tension.  The estimate that we performed of the $2
\to 2$ scattering amplitude is equivalent to this t--channel summation, with
two modifications: first, in approximating $s^{\alpha_G(t)} \simeq s^2$, we
neglected the shift $\alpha't$; however, since $-t = \xi M_s^2$ with $\xi \le
10^{-2}$, the corrections due to this shift is $O(10^{-3})$ and therefore
negligible.  The second modification is that together with the graviton, we
summed also over the $t$-channel KK excitations.  But this is, indeed, a basic
feature of these models, as we recall from the fact that it is actually
responsible for the crossover of the gravitational force from $1/r^2$ to
$1/r^{2+n}$ as $r$ decreases below $r_n$.  Note further that whatever the 
intermediate $s$-channel states are, our general arguments above show that they
cannot conspire to remove the $s$ growth of the cross section. 

Finally, we would like to address the issue of compositeness of the 
colliding particles, since hadrons are certainly composites and even quarks
and leptons might exhibit compositeness as length scales smaller than the
current limits of order (few \ $\times$ TeV)$^{-1}$.  
It is well known that because of the universality of the static
gravitational interaction, it is independent of the compositeness of the
objects involved, and only depends on their total masses, as $F=G_N m_1
m_2/r^2$.  It is interesting to observe that the asymptotic behavior 
of the cross section, $\sigma \propto s/M_6^4$ is also independent of the
possible (partonic or preonic) structure of the colliding particles.  To show 
this, we let the colliding particles labelled 1 and 2 consist, respectively, 
of $N_1$ and $N_2$ partons (preons).  We 
first note that for particles with masses satisfying $m_i^2 << s$ and
hence negligible for the present purposes, one can approximate the kinematics
by setting $m_i=0$, so that $s = 2p_1 \cdot p_2 = 4E_1 E_2$, where the
energies $E_1$ and $E_2$ are specified in the center-of-mass frame or any 
Lorentz frame obtained from it via a boost along the collision axis 
$\hat {\bf k}$.  Writing the energy of each particle as the sum of the 
energies of its constituents, 
\beq
E_1 = \sum_{i=1}^{N_1} E_{1i} \ , \quad E_2 = \sum_{i=1}^{N_2} E_{2i}
\label{ej}
\eeq
and substituting, we obtain, for the total cross section
\beq
\sigma_{12} = \kappa \frac{s}{M_6^4} = \frac{4\kappa E_1E_2}{M_6^4} = 
\frac{\kappa}{M_6^4}\sum_{i=1}^{N_1}\sum_{j=1}^{N_2} E_{1i}E_{2j} = 
\sum_{i=1}^{N_1}\sum_{j=1}^{N_2} \sigma_{ij} 
\label{sigmacomp}
\eeq 
where $\kappa=4\pi$ in eq. (\ref{sigmafinal}) but its precise value is not
important here.  That is, we write the particle-1 particle-2 cross section
$\sigma_{12}$ as the incoherent sum of the $N_1N_2$ parton-parton (preon-preon)
cross sections and find that, to this order, $\sigma_{12}$ is invariant under
composition.  This is to be contrasted with a $\sigma_{12} \propto s^2$ cross
section, which would not have this invariance.  In passing, we also note that
because of the specific limitation $-t << \xi M_s^2 << M_s^2$ that we used, and
the fact that the string scale is $M_s^2$, we do not expect
here any form-factor effects on the graviton+KK couplings.  Thus, the cross
section behavior $\sigma \propto s$ in eq. (\ref{sigmafinal}) can be
interpreted in a certain sense, e.g., in the context of composite models, as
preserving the counting of degrees of freedom. (The criterion of preserving
degrees of freedom and associated information content has also been used to 
infer a holographic principle in quantum gravity \cite{holo}.)

One may wonder whether the growth of the cross section over many decades of
accessible energies \cite{cn} does not constitute a fundamental difficulty and
potential fatal flaw, in principle, of the model.  If the truly naive estimate
$\sigma \propto s^2$ from the consideration of the $N$-graviton exchange
amplitude were valid, then it would appear that this would, indeed, be a
serious problem.  If, however, the cross sections grow only like $s$, as
suggested by our discussion above, then there may not be any fatal difficulty.
To see this point, let us assume that the forward dispersion relation for some
$2 \to 2$ scattering amplitude, such as $\nu + N \to \nu + N$ holds (where $N$
= nucleon) holds.  Then, with the optical theorem that the imaginary part of
the forward scattering amplitude $A(s,t=0)$ satisfies $Im(A(s,t=0)) \propto s
\sigma(s) \propto s^2$, it follows that in these models, this forward
dispersion relation would need one more subtraction than in the usual case
where $\sigma \le const. \times ln^2 s$.  The subtraction term is likely
to involve the factor $\alpha' s \sim s/M_s^2$, which is also what the
field-theory limit, via the Taylor-series expansion in $\alpha' s$, of string
theory scattering amplitudes would suggest.

\section{Effects of High-Energy Gravitational Scattering Cross Section} 

\subsection{Ultra-High Energy Cosmic Rays} 

In recent years, there has been considerable progress in the measurement of
ultra-high-energy UHE cosmic rays.  These have been observed with lab energies
up to $E \sim 3 \times 10^{20}$ eV \cite{uhecr}. The cosmic rays consist of
protons together with some nuclei having charge $Z > 1$ (the relative fractions
depending on the energy).  The highest-energy cosmic rays are of considerable
current interest because their energies are above the Greisen-Zatsepin-Kuzmin
(GZK) upper bound \cite{gzk} of $E_{GZK} \simeq 5 \times 10^{19}$ eV, which was
based on the fact that above this energy, the cosmic rays would lose energy by
scattering off the cosmic microwave background photons and producing pions, via
the reaction 
\beq 
p + \gamma_{CMB} \to N + \pi
\label{gzkreaction}
\eeq
which can proceed resonantly
through the $\Delta(1232)$.  A number of mechanisms have been proposed to try
to account for this \cite{uhecr}.  For the illustrative energy $E = 10^{19}$
eV, i.e., $s \simeq 2 \times 10^4$ TeV$^2$, the cross section estimate
(\ref{sigmafinal}) for the new graviton+KK contribution to $pp$ scattering,
together with the lower bound (\ref{m6bound}), gives $\sigma_{grav+KK} \lsim
10^{-34}$ cm$^2$.  This is quite small compared to the usual hadronic $pp$
cross section, which, at this energy, is $\sigma_{pp} \simeq 150$ mb \ $= 1.5
\times 10^{-25}$ cm$^2$.  Thus, using our unitarized estimate
(\ref{sigmafinal}), we find that the phenomenological impact of models with two
large compact dimensions on UHE $pp$ scattering is innocuous.

\subsection{Ultra-High Energy Neutrino Scattering} 

A striking feature of the new contribution to high-energy scattering cross
sections due to graviton+KK exchange, eq. (\ref{sigmafinal}), is the fact that
it is projectile-independent, and thus the same for the scattering of $pp$,
$\gamma p$, $\nu p$, etc.  We focus here on ultra-high energy neutrinos; one of
the reasons why these are of current interest is that they occur as the decay
products of the pions produced by the reaction (\ref{gzkreaction}) and
analogous reactions in which the protons scatter off of the radiation field of
a source such as a gamma-ray burster (GRB) \cite{nurev,waxman}.  Calculations
of UHE $\nu p$ and $\bar\nu p$ weak charged and neutral current cross sections
find that these grow with energy approximately as $E^{0.36}$ \cite{quigg}.
Since this is a lower power than the linear growth in $s$, and hence $E$,
exhibited by our estimate (\ref{sigmafinal}) of the cross section due to
graviton+KK exchange, the latter would eventually exceed the neutrino cross
sections based on $W$ and $Z$ exchange.  Using the lower bound (\ref{m6bound})
with our estimate (\ref{sigmafinal}), yields, for the contribution of
gravitational scattering to $\nu p$ (or $\bar\nu p$) scattering, the result
$\sigma_{grav+KK} \sim 10^{-44} (E_\nu /{\rm GeV})$ cm$^2$, so that at $E_\nu
\sim 10^{21}$ eV, this gravitational contribution is $\sigma_{grav+KK} \sim
10^{-32}$ cm$^2$, roughly 10 \% of the sum of the conventional charged and
neutral current $\nu p$ (or $\bar\nu p$) scattering cross sections, as
calculated in Ref. \cite{quigg}.  Clearly our estimate (\ref{sigmafinal}) and
the lower bounds (\ref{mbound}), (\ref{m6bound}) are only approximate, so the
implication that we draw from the present discussion is that in theories with
large compactification radii and TeV scale quantum gravity, it is possible for
gravitational scattering to make a non-negligible contribution to ultra-high
energy $\nu p$ and $\bar\nu p$ total scattering cross sections. If, indeed,
this new gravitational contribution to $\nu p$ and $\bar\nu p$ scattering is
significant, this will be important for large detectors like AMANDA
\cite{nurev} and the Auger Cosmic Ray Observatory \cite{auger} which will
measure these UHE neutrinos; if one just used the conventional CC + NC cross
sections, one would infer a somewhat larger (anti)neutrino flux than would
actually be present.  Since at energies $E \gsim 10^{15}$ eV, the interaction
length for (anti)neutrinos is smaller than the diameter of the earth, and for
$E \gsim 10^{18}$ eV, the earth is opaque to (anti)neutrinos \cite{quigg}, it
follows that this possible enhancement of the observed event rate would only
occur for downward going (anti)neutrinos.  It should be emphasized, however,
that our discussion concerning the gravitational cross section is subject to
the uncertainty regarding the characteristics of the dominant final states in
gravitational scattering of UHE $\nu p$ and $\bar\nu p$ collisions.  In
particular, it could be that these collisions involve considerable
multi-graviton-KK production, in which case it would be interesting also to
study the resultant propagation and interaction of these KK modes.

\section{Light Neutrinos}

For a number of years there has been accumulating evidence for neutrino masses
and associated lepton mixing; at present, it is generally considered that the
strongest evidence arises from the deficiency of the solar neutrino flux
measured by the Homestake, Kamiokande, SuperKamiokande, GALLEX, and SAGE
detectors \cite{solarnu} and the atmospheric neutrino anomaly first observed by
the Kamioka and IMB detectors, confirmed by the Soudan-2 detector, and recently
measured with high statistics by the SuperKamiokande detector \cite{atmos}.  An
appealing explanation of this data is neutrino oscillations and associated
neutrino masses and mixing.  As is well known, fits to the solar neutrino data
yield values of $\Delta m_{ij}^2 = |m(\nu_i)^2 - m(\nu_j)^2|$ of order
$10^{-5}$ eV$^2$ for $\nu_e \to \nu_j$ (although ``just-so'' oscillations
feature $\Delta_{1j} \sim 10^{-10}$ eV$^2$), and the SuperKamiokande experiment
fits its atmospheric neutrino anomaly as being due to the neutrino oscillations
$\nu_\mu \to \nu_\tau$ (or $\nu_\mu \to \nu_s$, where $\nu_s$ is a light
electroweak-singlet neutrino) with central value $\Delta_{23} \sim 2 \times
10^{-3}$ eV$^2$ \cite{atmos,lsnd}.  This data, then, suggests neutrino masses
in the range $ \lsim 0.1$ eV for the mass eigenstates $\nu_i$, $i=1,2,3$ whose
linear combinations comprise the neutrinos $\nu_e$, $\nu_\mu$, and $\nu_\tau$.
Many details require further experimental clarification, but the most striking
aspect of this is the smallness of these masses relative to the masses of the
quarks and charged leptons.  Indeed, independently of these suggestions for
nonzero neutrino masses, direct mass limits yield upper bounds on the masses of
the mass eigenstates comprising $\nu_\ell$, $\ell=e,\mu,\tau$ are much smaller
than the masses of the respective charged leptons.

A fundamental theory should therefore give an explanation of why neutrinos are
so light.  The well-known seesaw mechanism does this \cite{seesaw}.  However,
we are not aware of any mechanism of comparable simplicity and elegance in
theories in which the highest fundamental scale is of order 10 TeV.  To see the
problem, we recall how the seesaw mechanism works.  This is based on the
assumption that there exist some number of electroweak-singlet neutrinos,
$\{N_R \}$.  In SO(10) grand unified theory, this is not an {\it ad hoc}
hypothesis but an intrinsic feature; the fermions of each generation transform
according to the 16-dimensional spinor representation, $16_L = 10_L + 5^c_L +
1_L$, where the SU(5) representations are indicated, and the SU(5)-singlet
state is the $N_L^c$.  This also fixes the number of electroweak-singlet
neutrinos as equal to the number of the usual fermion generations, $N_{gen}=3$.
These electroweak-singlet singlet neutrinos lead to Dirac mass terms of the
form 
\beq 
\sum_{i,j} \bar \nu_{j,L} M_{D,ij} N_{i,R} + h.c. 
\label{dirac}
\eeq
and 
Majorana mass terms of the form $\sum_{i,j} N_{i,R}^T M_{R,ij} C N_{j,R} + 
h.c.$, where $i,j$ are generation indices.  Because
the Majorana terms are electroweak-singlet operators, the coefficients in the
matrix $M_R$ should not be related to the electroweak symmetry breaking
scale, but instead, taking a top-down view, would naturally be of order the
largest mass scale in the theory.  The specific SO(10) GUT provides an explicit
realization of this expectation, with the coefficients in $M_R$ naturally of
order the GUT scale. 
 If $M_R$ has maximal rank, then the diagonalization of the full
Dirac + Majorana neutrino mass matrix yields a set of heavy, mainly
weak-singlet Majorana neutrinos with masses of order $M_{GUT}$ and the three
light mass Majorana eigenstates, which are mainly linear combinations of the
observed isodoublet weak neutrino eigenstates.  The masses of the observed
left-handed neutrinos are determined, to leading order in $M_{ew}/M_R$, as the
eigenvalues of the matrix 
\beq 
M_{\nu} = -M_D M_R^{-1} M_D^T
\label{mnu}
\eeq 
For a generic matrix $M_R$ (with maximal rank), this provides a natural
explanation of why these masses are much smaller than the scale of the quarks
and charged leptons.  At a more phenomenological level, even without any
explicit consideration of electroweak-singlet neutrinos, one can also get 
naturally small neutrino masses by dimension-5 operators of the form 
$(1/M_X){\cal L}_L^T C {\cal L}_L \phi \phi$, where 
${\cal L} = (\nu_\ell, \ell)$ is the $\ell$'th lepton doublet and $\phi$
is the $Y=1$ Higgs in the standard model
or its supersymmetric extension (for an explicit expression, see, e.g.,
Ref. \cite{pdg96}).  Such higher-dimension operators occur naturally if one 
regards the standard model as an effective low-energy field theory; however,
the success of this model requires that one make $M_X >> M_{ew}$.  The
resulting neutrino masses are $\sim M_{ew}^2/M_X$ and hence are naturally
small. 

In a theory in which there are no fundamental mass scales $>> M_{ew}$, 
this type of mechanism is not available to explain small neutrino masses.  
It is true that rapid running of gauge and Yukawa couplings $g_a$ and 
$Y_{ij}$ will occur once the energy gets to the point where the standard 
model fields feel the extra dimensions, since $dim(g) = dim(Y) = (4-d)/2$ and
hence are dimensional \cite{add2}.  However, the corresponding powers of $E$ in
the effective couplings divide out in ratios, so it is not clear how this
mechanism would specifically pick out the neutrinos to be so light. 

Of course, one notices that for the case $n=2$, with $M_6$ near its lower bound
(\ref{m6bound}), the theory exhibits a mass scale $r_2^{-1} \sim 0.07$ eV, as
in eq. (\ref{r2}), and the square of this scale, $r_2^{-2} \sim 5 \times
10^{-3}$, is comparable to the value of $\Delta_{23}$ inferred by the
SuperKamiokande collaboration to fit their atmospheric anomaly.  But it is not
clear to us how to relate the compactification mass scale $r_2^{-1}$ to masses
of fermions, and in particular, neutrinos.  It is also not clear how to get a
generational (family) structure, and corresponding spectrum of mass values, out
of a single $r_2$, although one should remember that the assumption of a single
compactification radius was only made as a simplication, and, in the absence of
some symmetry to guarantee equal compactification radii for the different extra
dimensions, the generic expectation is that these radii would be at least
somewhat different.

\section{Quark-Lepton Unification}

One of the deepest goals of a physical theory is to provide a unified
understanding of a wide range of different quantities and phenomena.  Thus, one
of the greatest appeals of grand unified theories is that they provide (i) a
basic unification of the QCD and electroweak gauge interactions, embedding the
associated SU(3) and SU(2) $\times$ U(1)$_Y$ gauge groups in a simple group;
and (ii) correspondingly, a unification of quarks and leptons, since these are
transformed into each other (or their conjugates) by the GUT gauge
transformations.  Proton and bound neutron decay is a natural consequence of
this unification, and the slow, logarithmic running of the gauge couplings is
nicely concordant with the high mass scale $M_{GUT}$ which is necessary to
suppress nucleon decay adequately to agree with experimental limits.
Consequently, a fundamental concern with theories whose largest fundamental
scale is of order 10 TeV, say, is that this scale may be so close to the
electroweak scale, where the standard model gauge group is definitely not
unified in a larger group, that it may be unnatural for such a unification to
occur.  Even if the rapid power-law running of gauge couplings were to lead to
some sort of unification of these couplings, one would obviously not want this
to be accompanied by the onset of a simple grand unified group with GUT scale
of order 10 TeV, since this would generically lead to prohibitively rapid
proton decay.  This is not to say that one cannot contrive special mechanisms
to suppress proton decay in such models (see, e.g., \cite{add2,ddg,st}), but
true quark-lepton unification generically leads to proton decay at the scale of
unification, so if one attempts such unification at the TeV scale, one must
resort to special suppression mechanisms.

\section{Equivalence Principle}

A basic experimental property of gravity is the equivalence principle, that
inertial and gravitational mass are equal, or, in Einstein's formulation, the
statement that the effects of a uniform gravitational field are equivalent to
the effects of a uniform acceleration of the coordinate frame and, related to
this, that gravity couples only to mass, independent of other quantum numbers
of the particles involved. This has been tested in laboratory and solar system
measurements down to the level of about $10^{-12}$ \cite{will,wbook,damour}.
This principle can be derived from the properties of the coupling of the
massless spin-2 graviton required to yield Lorentz-invariant amplitudes
\cite{w65,wbook}.  Thus, modifications to this formalism will, in general, lead
to violations of the equivalence principle.  For example, it was recognized
that such violations would result from a massless or sufficiently light ($m
\lsim 10^{-3}$ eV) dilaton in string theories of quantum gravity
\cite{dil}.  Moreover, one might question how natural it is to propose
that the fundamental scale of gravity is close to the scale of an interaction,
namely the electroweak interaction, which does not couple to particles in a
manner dictated by the principle of equivalence but instead according to their
gauge group representations.  We obtain a rough estimate of the violation of
the equivalence principle in the present class of theories as follows.  As
discussed before, the KK modes of the gravitons couple like the graviton, but,
from a four-dimensional point of view, have a mass given by (\ref{mukk}).  The
minimal value of this mass is just $\mu = r_n^{-1}$ and, in the $n=2$ case
where $r_n$ is largest, this is listed in eq. (\ref{r2}) for $M_6=30$ TeV as
$\mu \sim 0.07$ eV.  The long-distance effects of this lightest KK particle are
suppressed by a usual exponential, $e^{-\mu r}$.  For values of $r$ in the
macroscopic range, say $r \sim 10$ cm, this suppression factor is far below the
$10^{-12}$ level.

\section{Cosmological Constant}

The issue of the cosmological constant $\Lambda$ and the associated vacuum
energy density 
\beq
\rho_\Lambda = \frac{\Lambda}{8 \pi G_N}
\label{rholambda}
\eeq
has been a longstanding and vexing one in cosmology.  As is well known, vacuum
fluctuations of quantum fields naturally yield contributions to $\rho_\Lambda$ 
that are many orders of magnitude larger than the upper limit
$\Omega_\Lambda \lsim 2$ derived from the requirement not to overclose the
universe, where 
\beq
\Omega_\Lambda = \frac{\rho_\Lambda}{\rho_c} = \frac{\Lambda^2}{3H_0^2}
\label{omegalambda}
\eeq
and the critical mean mass density, $\rho_c$, is given by 
\beq 
\rho_c = \frac{3H_0^2}{8\pi G_N} = 1.05 \times 10^{-5} \ h_0^2 \ \ {\rm 
GeV/cm}^3 = 3.97 \times 10^{-11} \biggl (
\frac{h_0}{0.7} \biggr )^2 \ {\rm eV}^4
\label{rhocrit}
\eeq 
with $H_0$ the current value of the Hubble constant $H$ defined as
$H=R^{-1}dR/dt$, where $R(t)$ is the scale factor in the standard
Friedmann-Robertson-Walker metric.  In terms of $h_0 = H_0/(100 \ {\rm
km/sec/Mpc})$, the current value of the Hubble constant is $0.5 < h_0 < 0.85$
(e.g. Ref. \cite{pdg}).  Specifically, in quantum gravity without supersymmetry
one would generate a contribution to $\rho_\Lambda$ of order $M_{Pl}^4$, a
factor of $10^{125}$ times larger than the upper limit.  Even if this is
avoided by supersymmetry, the electroweak symmetry breaking contributes a term
of order $M_{ew}^4$ which is $10^{65}$ times larger than $\rho_c$, and this
cannot be avoided by supersymmetry, since supersymmetry must be broken at
energies of at least the scale $M_{ew}$.  From a field-theory point of
view, one would argue that since no symmetry prevents a nonzero $\Lambda$, it
should be present \cite{wl}.  Indeed, recent data on type Ia supernovae
\cite{berk,harv} and other observational data \cite{lambdagen} suggests
that $\Omega_\Lambda$ is nonzero and of order unity (see, e.g., Figs. 6 and 7
of Ref. \cite{harv}). 

It is natural that whenever a new type of theory is proposed, one tries to
explore its implications for the cosmological constant.  In any theory with
compactified dimensions, one is led to investigate the effects of
compactification on vacuum fluctuations and resultant contributions to the
cosmological constant \cite{kkfluct}.  One observation is that if one simply
extracts mass and length scales phenomenologically from $\rho_\Lambda$ by using
the value $\Omega_\Lambda \sim 1$ suggested by the supernovae data
\cite{harv,berk}, setting 
\beq
\rho_\Lambda \equiv r_\Lambda^{-4}
\label{rlambda}
\eeq
then one
obtains
\beq
r_\Lambda^{-1} \sim 2.5 \times 10^{-3} \ {\rm eV}, \ i.e., \ 
r_\Lambda \sim 0.8 \times 10^{-2} \ {\rm cm}
\label{rlambdavalue}
\eeq 
This is intriguingly close to the value of the compactification radius
$r_2 \sim O(1)$ millimeter implied by eq. (\ref{rnvalue}) if one were to use
the original estimate $M_{4+n}=M_6 \sim 1$ TeV for the $n=2$ case.  The
supernova cooling constraint of $M_6 \gsim 30$ TeV in eq. (\ref{m6bound})
reduces $r_2$ to $\sim$ a micron rather than $\sim$ a millimeter, thereby
removing a close similarity to $r_\Lambda$.  Nevertheless, in view of the
disparity of 65 orders of magnitude between $M_{ew}^4$ and the upper bound on
$\rho_\Lambda$, there is perhaps a hint of some interesting physics relating
$r_n$ and $r_\Lambda$.  In the theoretical context of both KK theories and
D-branes, one is also led to consider Casimir-type contributions to the vacuum
energy density \cite{cas}, which are of the form $U_{Cas.} \propto r^{-4}$,
where $r$ is a relevant distance scale, such as the compactification scale.  In
this context, there is, indeed, a natural connection between $\rho_\Lambda$ and
$r_n^{-4}$ (see also Refs. \cite{st,sundrum,stab}).

\section{Conclusions}

Theories with large compactification radii and a very low string scale
represent a radical departure from previous fundamental particle theory
including grand unification and conventional string theory.  They serve to
emphasize that a vast 33 orders of magnitude separate the length scale of about
a centimeter at which gravity has been experimentally measured from the Planck
length of $10^{-33}$ cm, and it is quite possible that new phenomena occur in
these 33 decades of energy or length which significantly modify gravity from
one's naive extrapolations.  We believe that it is healthy to formulate and
study such radical challenges to the orthodox paradigm in order to assess how
experimentally viable they are.  Indeed, an intense effort to analyze the
phenomenology of these types of theories is currently underway.  In these
theories the Planck mass becomes a secondary rather than basic quantity, and is
expressed via eq. (\ref{rnvalue}) as a function of the short-distance string
scale and the compactification radius.  In this paper we have contributed a few
results to this effort.  We have made an estimate of high-energy gravitational
scattering cross sections and have discussed some of the implications for
ultra-high-energy cosmic ray and neutrino scattering.  We have also commented
on some other topics, including naturally light neutrino masses, 
quark-lepton unification, the equivalence principle, and the
cosmological constant.  Many more interesting questions deserve study.

\vspace{10mm}

We are grateful to Fred Goldhaber, Frank Paige, Bill Weisberger, and especially
to Gary Shiu for sharing with us his expertise on this subject and
teaching us about his work with Henry Tye, Ref. \cite{st}.  S. N. thanks the
Israel-U.S. Bi-National Science Foundation.  The research of R. S. was
partially supported by the U.S. National Science Foundation under the grant
PHY-97-22101.

\vfill
\eject


\begin{thebibliography}{99}
\bibitem{dbrane}{Here a D$_p$-brane is an extended object with $p$ spatial
dimensions, on which a string ends.  For the usual superstring
$X(\sigma,\tau)^\mu$ in $d=10$ spacetime, the boundary conditions for a single
D$_p$-brane can be taken to be 
$\partial_\sigma X^\mu=0$ (Neumann) for $\mu=0,1,...,p$ and 
$X^\mu=0$, $\mu=p+1,...,9$ (Dirichlet).}
\bibitem{reggecuts}{We recall from the original application of string theories
to dual resonance models of hadron scattering, that there are difficulties 
with iterated exchanges of Regge trajectories that have intercepts 
$\alpha(t=0)=2$ and higher as the relevant Regge cuts (at $2\alpha(0)-1$) will
lie even higher.  Until a complete unitarization program is carried out, one
would not be able to derive rigorously the true symptotic behavior of the cross
section.}
\bibitem{cn}{It has been argued \cite{cnpaper} that energies of order the usual
Planck mass $M_{Pl} \sim 10^{19}$ GeV may be unreachable.  We still have, even
in this case, eight orders of magnitude separating the ultra-high
energy cosmic rays with $E \sim 10^{11}$ GeV and the above Planck energy.}
\end{thebibliography}
\end{document}